\documentclass[aps,twocolumn]{revtex4-1}
\usepackage{graphicx}
\usepackage[justification=justified,width=\linewidth]{caption}
\usepackage{subcaption}
\usepackage{amsmath}
\usepackage{amsfonts}
\usepackage{amsthm}
\usepackage{amssymb}
\usepackage{amsbsy}
\usepackage{wasysym}
\usepackage{bm}
\usepackage{mathrsfs}
\usepackage{color}
\usepackage{times}
\usepackage[resetlabels]{multibib}

\begin{document}

\title{Phase separation kinetics of segregating fluid mixtures in the presence of quenched disorder}
\author {Rounak Bhattacharyya and Bhaskar Sen Gupta}
\affiliation{Department of Physics, School of Advanced Sciences, Vellore Institute of Technology, Vellore, Tamil Nadu - 632014, India}

\date{\today}
\begin{abstract} 
Quenched or frozen-in structural disorder is ubiquitous in real experimental systems. Much of the progress is achieved in understanding the phase separation of such systems using the diffusion-driven coarsening in Ising model with quenched disorder. But there is a paucity of research in the phase-separation kinetics in fluids with quenched disorder. In this paper, we present results from a detailed Molecular dynamics simulation, the effects of randomly placed localized impurities on the phase separating kinetics of binary fluid mixture. Two different models are offered for representing the impurities. We observe a dramatic slowing down in the pattern formation with increasing impurity concentration. This sluggish domain growth kinetics follows  power-law with a disorder-dependent exponent. The correlation function and structure factor show a non-Porod behavior, indicating the roughening of domain interfaces. We have also studied the effect of quenched disorder on the aging dynamics by calculating the two-time order parameter auto correlation function and find that the Fisher and Huse scaling law holds good in presence of quenched disorder.
\end{abstract}
\maketitle

\section{Introduction} 
Phase separation, the spontaneous segregation of a homogeneous mixtures into multiple distinct phases, is a ubiquitous phenomena observed in various systems, ranging from colloidal suspensions and polymers to liquid 
crystals and biological fluids \cite{Fisher,Stanley,Binder, Jones, Bray}. In the context of binary fluid mixtures, the phase 
separation process involves the demixing of two immiscible components, resulting in the formation of distinct liquid domains or phases. Understanding the kinetics of phase separation is crucial for predicting and controlling the final 
composition, morphology, and stability of the resulting phases, both from a scientific perspective and in terms of industrial applications \cite{chen,worrell,nomura}.

 When a homogeneous multicomponent system is rapidly cooled within the miscibility gap, various ordered phases emerge and expand over time. For our particular interest, in the evolution of a binary mixture (composed of substances A and B), distinct phases with a higher concentration of either A or B are formed. This non-equilibrium phenomenon has been extensively explored through analytical methods, experimental investigations, and numerical simulations \cite{Binder1,Siggia,Furukawa,Miguel,Tanaka,Beysens,Tanaka1,Kendon, Puri, Dutt,Laradji, Thakre, Ahmad}.

In the study of phase ordering dynamics in fluid mixtures, the growth of ordered phases in isotropic systems is typically described by a time-dependent length scale, denoted as $\ell(t)$, where $t$ represents the time after a rapid cooling \cite{Bray}. It is now well understood that the correlation function of the order parameter, denoted as $C(\vec{r}, t)$, and its Fourier transform $S(\vec{k}, t)$, exhibit scaling phenomena. Here, $\vec{r}$ represents the distance between two spatial points, and $\psi(\vec{r}, t)$ is an order parameter related to the local density difference between particles A and B.

The scaling behavior of the correlation function is described by the spatial correlation scaling equation \cite{Binder2}
\begin{equation}
C(r, t) \equiv f[r/\ell(t)].
\end{equation}

Similarly, the scaling behavior of the structure factor is given by
\begin{equation}
S(k, t) \equiv \ell^d f[k\ell(t)]
\end{equation}
where $d$ is the dimension of the system and $k$ is the magnitude of the wave vector. The average domain size $\ell(t)$, appearing in both equations, generally follows a power law
\begin{equation}
\ell(t) \sim t^\alpha
\end{equation}
The value of $\alpha$ depends on the dominant transport mechanism during the phase separation process. For example, in binary alloys, diffusion dominates, leading to $\ell(t) \sim t^{1/3}$ for systems with dimension $d \geq 2$. This is known as the Lifshitz-Slyozov (LS) growth law \cite{Bray, Binder3}. In the case of binary fluids, hydrodynamics plays a major role, resulting in different values of $\alpha$ depending on the dimensionality and regime: $\alpha = 1/3 \rightarrow 1/2 \rightarrow 2/3$ for $d = 2$ and $\alpha = 1/3 \rightarrow 1 \rightarrow 2/3$ for $d = 3$. Both diffusive growth ($\alpha = 1/3$) and viscous hydrodynamic growth ($\alpha = 1$) have been observed in numerous numerical simulations, while the inertial hydrodynamic regime ($\alpha = 2/3$) has only been observed in lattice-Boltzmann simulations \cite{Siggia, Furukawa}.

Studying the two-time correlations $C_{\psi\psi}(r,t, t_w)$, where $t$ is the observation time and $t_w$ is the waiting time, is a useful tool for analyzing non-equilibrium dynamics. In equilibrium systems, the two-time correlation function exhibits time translation invariance, meaning that the results overlap due to the time displacement $t - t_w$. However, in nonequilibrium systems, time translation invariance breaks down, and with increasing age ($t_w$) of the system, the decay of $C_{\psi\psi}(r,t, t_w)$ becomes slower. In the diffusive regime, $C_{\psi\psi}(r,t, t_w)$ follows a scaling behavior \cite{FisherHuse,Desai} 
\begin{equation}
C_{\psi\psi}(r,t, t_w) \equiv x^{-\lambda}
\end{equation}
where $x = \ell/\ell_w$ and $\lambda$ is the Fisher-Huse exponent. In the viscous regime of phase ordering fluid mixtures, there is a distinct transition from a power-law decay to exponential behavior \cite{Ahmad2012}.

From the above discussions we can convincingly conclude that the kinetics of phase separation in pure systems has been well researched and comprehended. However, real systems are not free from impurities, which adds an additional layer of complexity to the subject. In experimental systems, we typically encounter two types of disorder: annealed or mobile impurities, and quenched or immovable impurities. The presence of disorder significantly impacts the dynamics and configuration of domains in a complex manner. A detailed numerical investigation of the effects of annealed disorder on the coarsening dynamics of a phase-separating binary fluid mixture can be found in Ref. \cite{RBBSG}. On the other hand, numerous studies have been conducted to understand diffusion-driven coarsening in Ising systems with quenched disorder using models like the random exchange Ising model (REIM) \cite{Puriparekh} and random field Ising model (RFIM) \cite{Chakrabarti1,Chakrabarti2}, considering both conserved and non-conserved order parameter dynamics. Quenched disorder acts as a background random potential for fluctuating degrees of freedom. In Ising systems, the sites with quenched disorder trap the domain boundaries. These disorder-induced traps become significantly influential at a specific length scale determined by the strength of the disorder. When a domain wall becomes trapped in a metastable state, its movement can only occur through thermal activation, overcoming the associated energy barrier. As a result, thermal fluctuations play a vital role in driving the long-term growth of domains in disordered Ising systems, in contrast to pure systems where thermal fluctuations have a minimal impact \cite{Puriparekh,Chakrabarti1,Chakrabarti2,Puri1, Huse, Grest, Srolovitz,Puri2,Bray2,Paul, Henkel,Aron,Brochard,Gennes1,Maher, Goh,Wiltzius}.

However, the understanding of phase-separation kinetics in fluid systems with quenched disorder is still in its early stages. This topic is important because the presence of disorder introduces additional spatial and temporal heterogeneities, and a complex morphological pattern and dynamics during phase separation is anticipated, which remains unexplored. In particular, to the best of our knowledge, there is no reported work involving the atomistic level simulation on the kinetics of phase separation of segregating liquid mixture with quench disorder. 

In this work, we perform an extensive numerical study to investigate the domain growth and aging dynamics of an immiscible symmetric binary fluid mixture in the presence of quenched disorder. This is of significant technological interest in industries and oil recovery processes. Here, we offer two different models for the choice of nature of the disorder. In the next section we present the two models used in this paper. The results of the domain growth dynamics and the aging behavior are presented in Sec. III. Finally, in Sec. IV we offer a summary and a discussion.

\section{Numerical Model and Method}
\subsection{Basic model for binary liquid} 
The numerical model we consider is a cubic box of volume $V$ containing a binary mixture of A and B types of particles having a 50:50 ratio. The average density of the system is $\rho=N/V=1$, where $N$ is the total number of particles. The interaction between two particles, labeled as $i$ and $j$, depends on their scalar distance $r=|\vec r_i-\vec r_j|$ and is described by the Lennard Jones (LJ) pair potential given by:

\begin{equation}
	\label{eq:lj_potential}
	U(r)=4\epsilon\left[\left(\frac{\sigma}{ r}\right)^{12}- \left(\frac{\sigma}{r}\right)^6\right].
\end{equation}
Here, $\epsilon$ represents the interaction strength between the particles having diameter $\sigma$. To improve computational efficiency, the range of interaction is truncated at $r=r_c=2.5\sigma$. The discontinuity in the potential and the corresponding force due to the insertion of this cut off is resolved by modifying the potential as follows, 
\begin{equation}
	\label{eq:mod_potential}
	u(r)=U(r)-U(r_c)-(r-r_c)\left(\frac{dU}{dr}\right)|_{r=r_c} 
\end{equation}

The parameters in the LJ potential are chosen as follows: $\sigma_{AA} = \sigma_{BB} = \sigma_{AB} = \sigma = 1.0$, and $\epsilon_{AA} = \epsilon_{BB} = 2\epsilon_{AB} = 1.0$. These parameter choices ensure that the phase ordering dynamics are energetically favorable, and the critical point occurs at $T_c = 1.42$, $k_B$ being the Boltzmann constant \cite{Das}. Here temperature is measured in units of $\epsilon/k_B$. This critical point is significantly different from the possible critical points of the liquid-solid and gas-liquid transitions. For numerical simplicity, we assume the values of $k_B$ and the mass of each particle to be unity. Additionally, the periodic boundary condition is applied in all three spatial directions to mitigate the finite size effect.

\subsection{Modeling quenched disorder in binary fluid}
In this study, the effect of quenched disorder is incorporated into the system by introducing a third type of particles. These particles are generated by randomly choosing a small fraction of A and B type of particles of equal number and denoting them as P type. Therefore, the mixture comprises A, B, and P types of particles. Various options exist when it comes to choosing the energy parameters $\epsilon_{PP}$, $\epsilon_{PA}$, $\epsilon_{PB}$ and the corresponding interaction length scales for the P type particles, such as incorporating multibody interactions or angle-dependent interactions. In our current analysis, for the sake of specificity, we opt to use identical energy and length-scale parameters for P type as for particle types A and B. Our emphasis lies in comprehending the impact of a specific selection of fixed disorder on the kinetics of phase separation, rather than attempting to emulate a particular experimental system. Therefore, the sole alteration made is that the mobility of the P particles is reduced to zero. As a result, they act as an impediment and influences the dynamics of the whole system. 

The P particles interact with the rest of the system via the  LJ potential with the parameters given as: $\epsilon_{PA} = \epsilon_{PB} = 0.5$. The potential is truncated at $r_c = 2^\frac{1}{6}\sigma$ to exclude any attractive force with A and B. Previous studies on systems with quenched disorder, particularly in the Ising model, have focused on varying the field strength or bond strength \cite{Chakrabarti1,Chakrabarti2}. In this study, the model of quenched disorder represents a basic form that closely resembles a realistic fluid system. We construct two different models to incorporate quenched disorder into the system, which are as follows:

\textbf{Model I:} In the first model, we randomly place the P type of particles across the system and freeze them at their respective positions throughout the simulation. Therefore, these immobile particles serve as quenched disorder. 

\textbf{Model II:} Here, initially we consider an interaction among the P type particles via the LJ potential with a cutoff of $r_c = 2.5\sigma$. This allows them to form clusters. The size of the clusters depends on the duration $\tau$ over which the interaction force is on. Finally, after a chosen $\tau$, we freeze the P particles, which consequently act as quenched disorder. 

\subsection{Phase separation dynamics} 
To study the effect of quenched disorder on the coarsening dynamics, we resort to molecular dynamics (MD) simulations in the NVT ensemble. The system consists of a total  N = 32768 particles. Velocity-Verlet algorithm \cite{Verlet} is used to integrate the equation of motion for each particle that evaluates the position and velocity during the MD simulation with the time step $\Delta t$ = 0.005. Here time is referred in unit of $\tau = (m\sigma^2/\epsilon)^{1/2}$. During the MD simulation, hydrodynamics preserving Nose Hover thermostat is used \cite{Nose}.  The density of the pinned particles $\rho_P$ is chosen to be significantly smaller than the average density of the system $\rho$ to preserve the original composition of the binary mixture. The simulation is performed with five different choice of pinning concentrations. The ensemble average of all the statistical quantities is computed from 50 separate simulations initiated with entirely distinct initial configurations.

The simulation is started with the preparation of an equilibrated homogeneous system by using MD simulation at high temperature $T=10.0$. The system is then quenched to $T = 0.77T_c$ at time $t=0$. In model \textbf{I}, the system is permitted to undergo evolution towards the thermodynamically favorable state until full phase separation is attained. The simulation is performed with five different choice of disorder concentration, $\rho_P=0.01, 0.02, 0.03, 0.04$ and 0.05. Throughout this process, the particles of type P are effectively immobilized by excluding them from participating in the dynamics.  

 In model \textbf{II}, after the quench at $t=0$, the system is permitted to undergo a phase separation process for a time duration of $\tau$. As a off critical composition, the minority phase (P type) domains form droplets. At this point, we freeze the P types of particles for the rest of the simulation and the domains formed by them are regarded as quenched disorder. On the other hand, the particles belonging to the majority phases are randomly reassigned as A and B types, maintaining their number ratio $50:50$. The system is then heated up again to T = 10.0 for equilibration to annihilate any memory effect. Finally, the system is quenched to $T = 0.77T_c$ at $t = 0$ and is allowed to evolve till the system is completely phase separated. We have explored the growth dynamics for three different choices of $\tau=50, 200$ and 800 at a fixed chosen disorder concentration $\rho_P=0.05$.
 
 \section{Results and discussion} 
\subsection{Domain morphology and growth dynamics}
 \begin{figure}
 	\centering
 	\begin{subfigure}[h!]{0.45\columnwidth}
 		\includegraphics[width=\columnwidth]{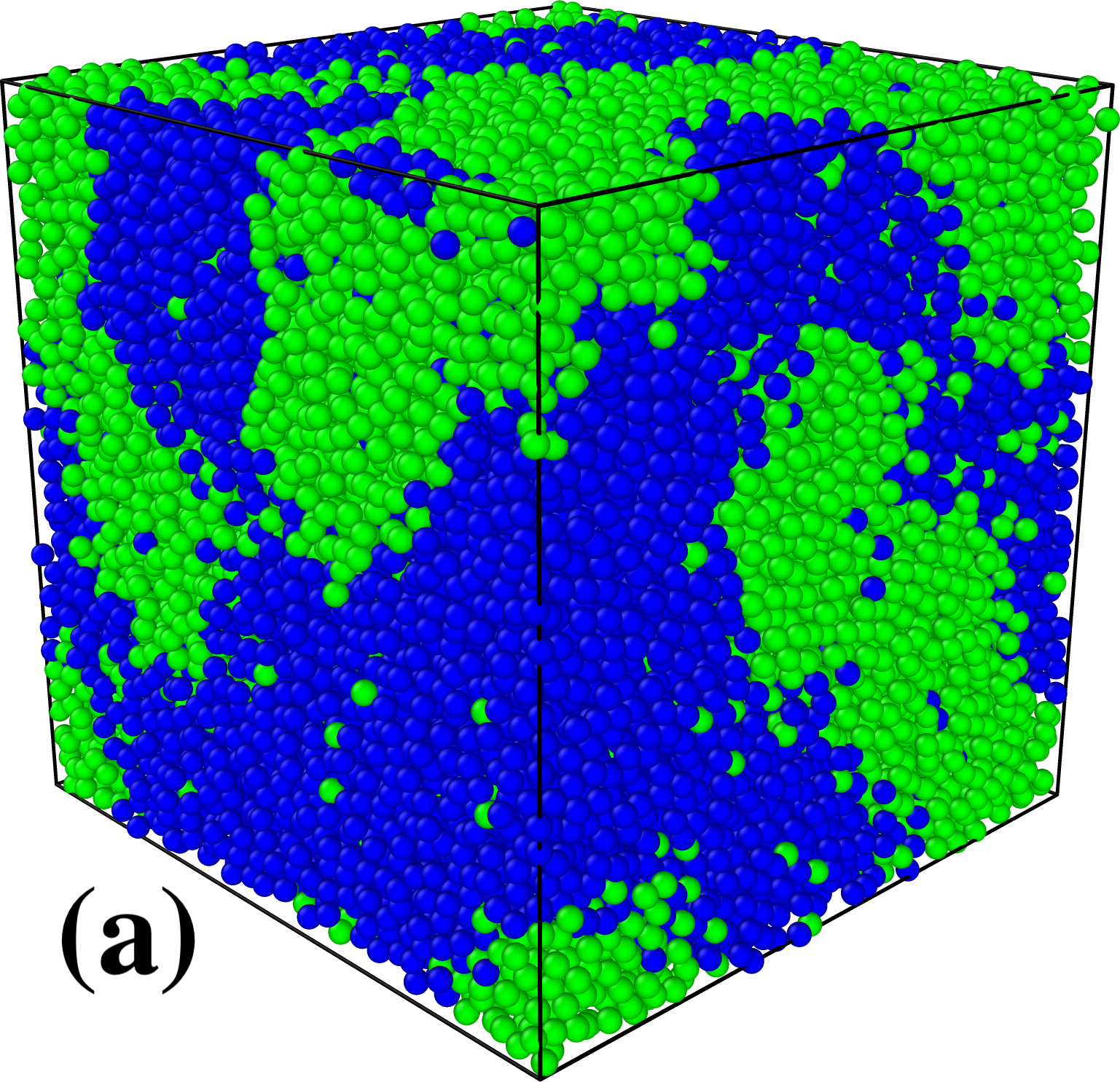}
 	\end{subfigure}
 \begin{subfigure}[h!]{0.45\columnwidth}
 	\includegraphics[width=\columnwidth]{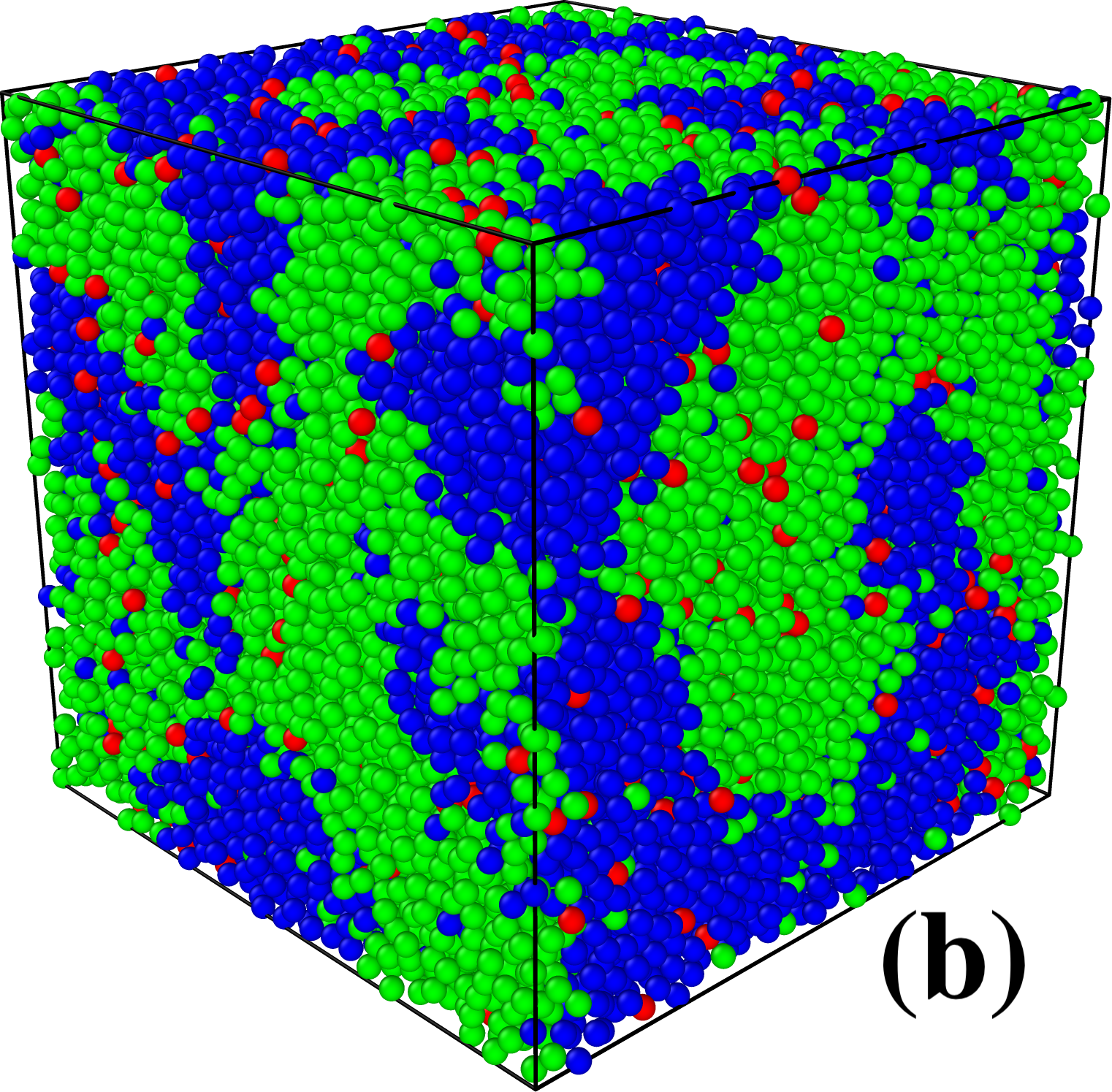}
 \end{subfigure}\\
	\begin{subfigure}[h!]{.4\columnwidth}
 		\centering
 		\includegraphics[width=\columnwidth]{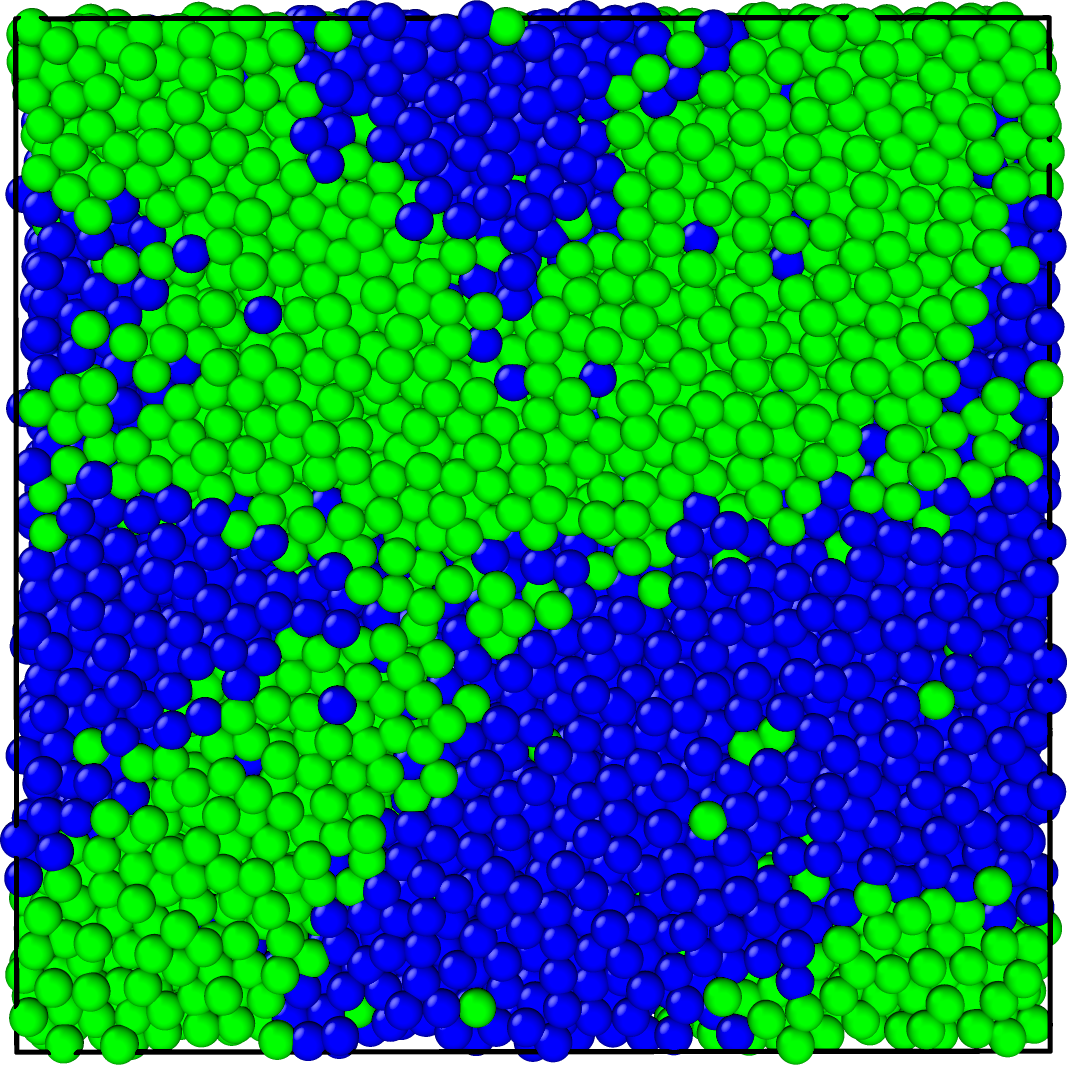}
 	\end{subfigure}
 \begin{subfigure}[h!]{.45\columnwidth}
 	\centering
 	\includegraphics[width=\columnwidth]{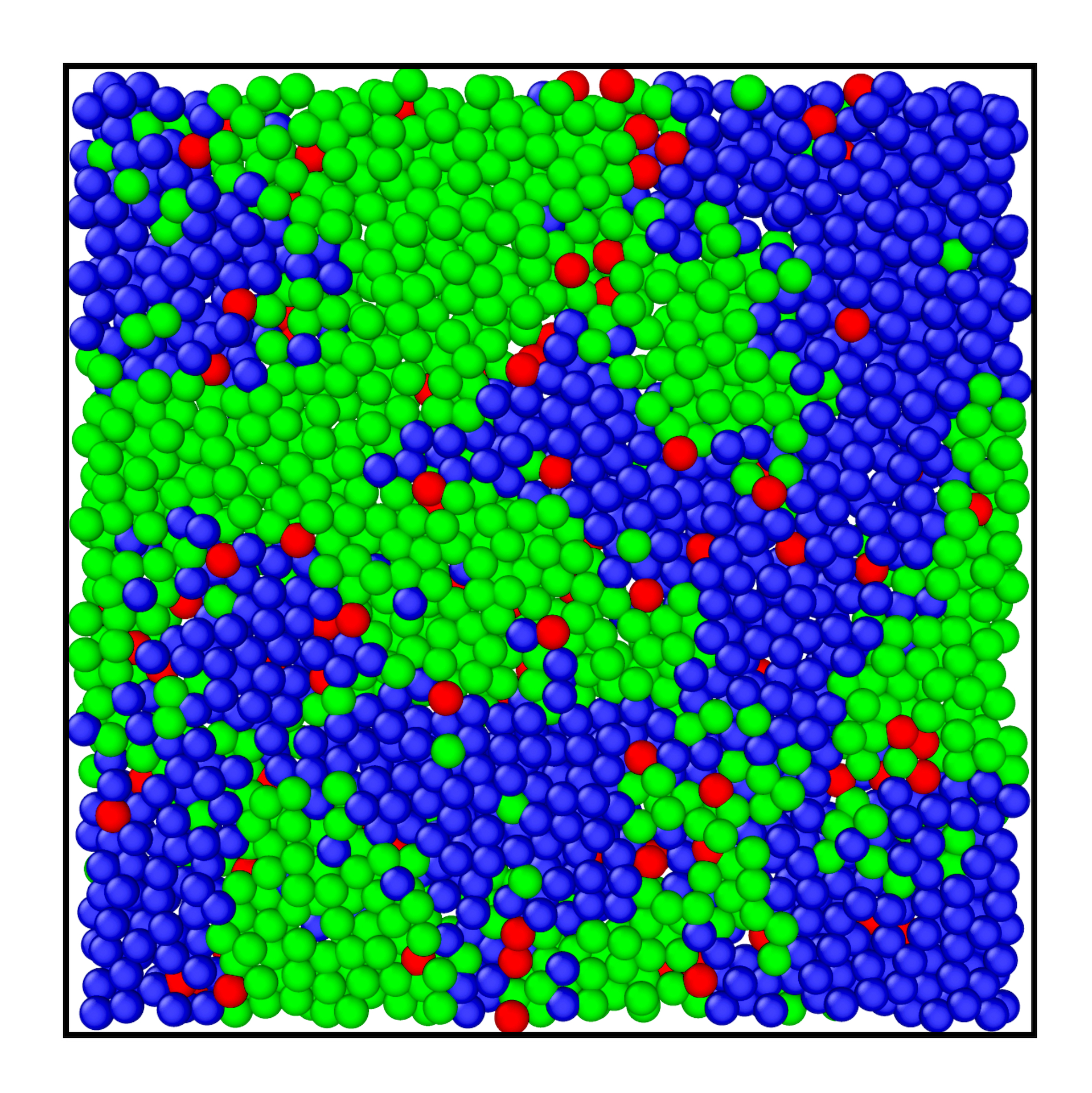}
 \end{subfigure}
 	\caption{Three-dimensional snapshots of our binary system at time $t=3000$ for $\rho_P=0.0$ and 0.05 are shown in (a) and (b) respectively for model \textbf{I}. The two dimensional cross sectional area of the same are shown in the lower panel. The A, B and P type particles are marked as green, blue and red color respectively.}
 	\label{fig1-snap}
 \end{figure}
 
 \begin{figure}
 	\centering
 	\begin{subfigure}[h!]{0.48\columnwidth}
 		\includegraphics[width=\columnwidth]{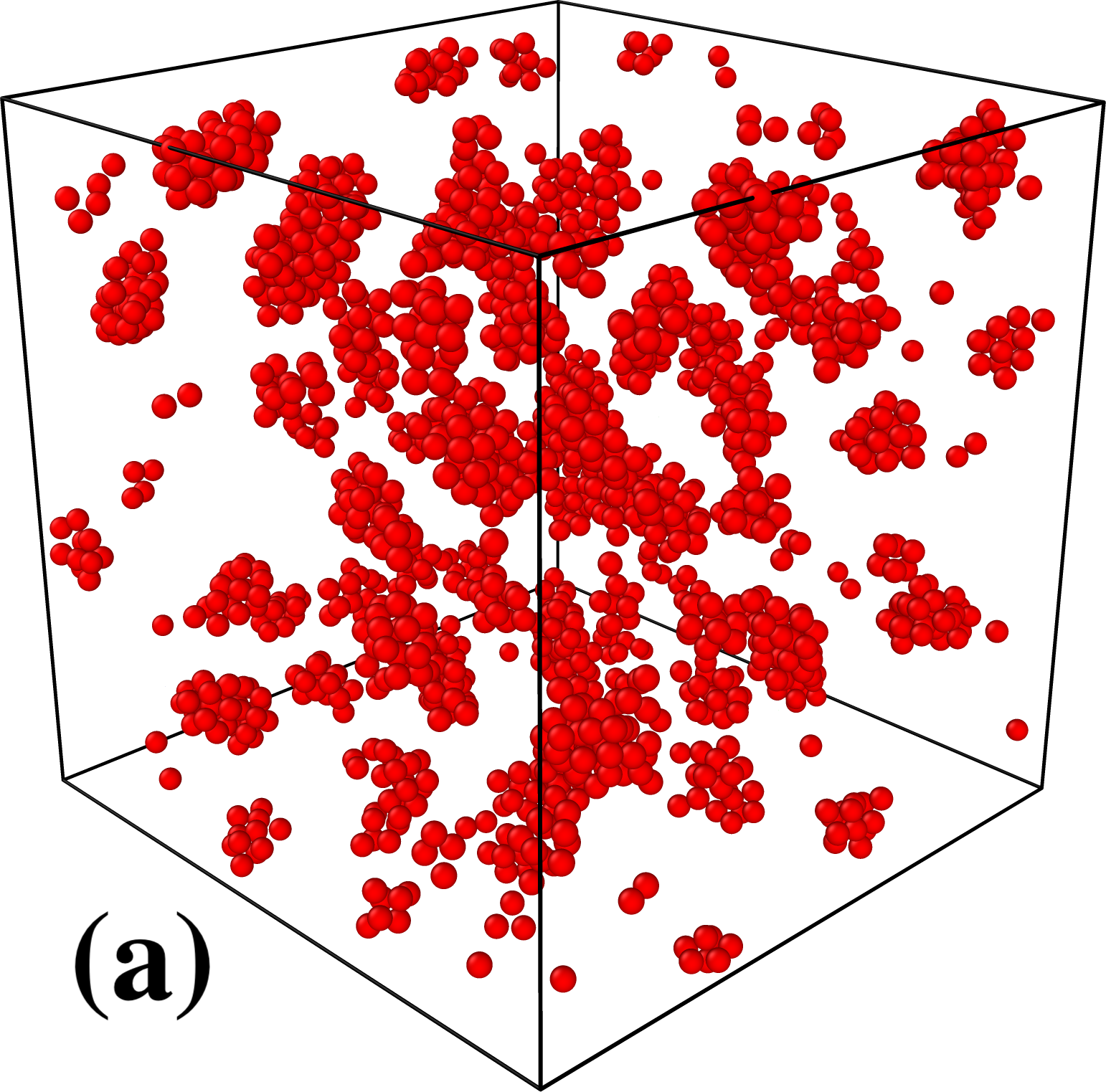}
 	\end{subfigure}
 \begin{subfigure}[h!]{0.48\columnwidth}
 	\includegraphics[width=\columnwidth]{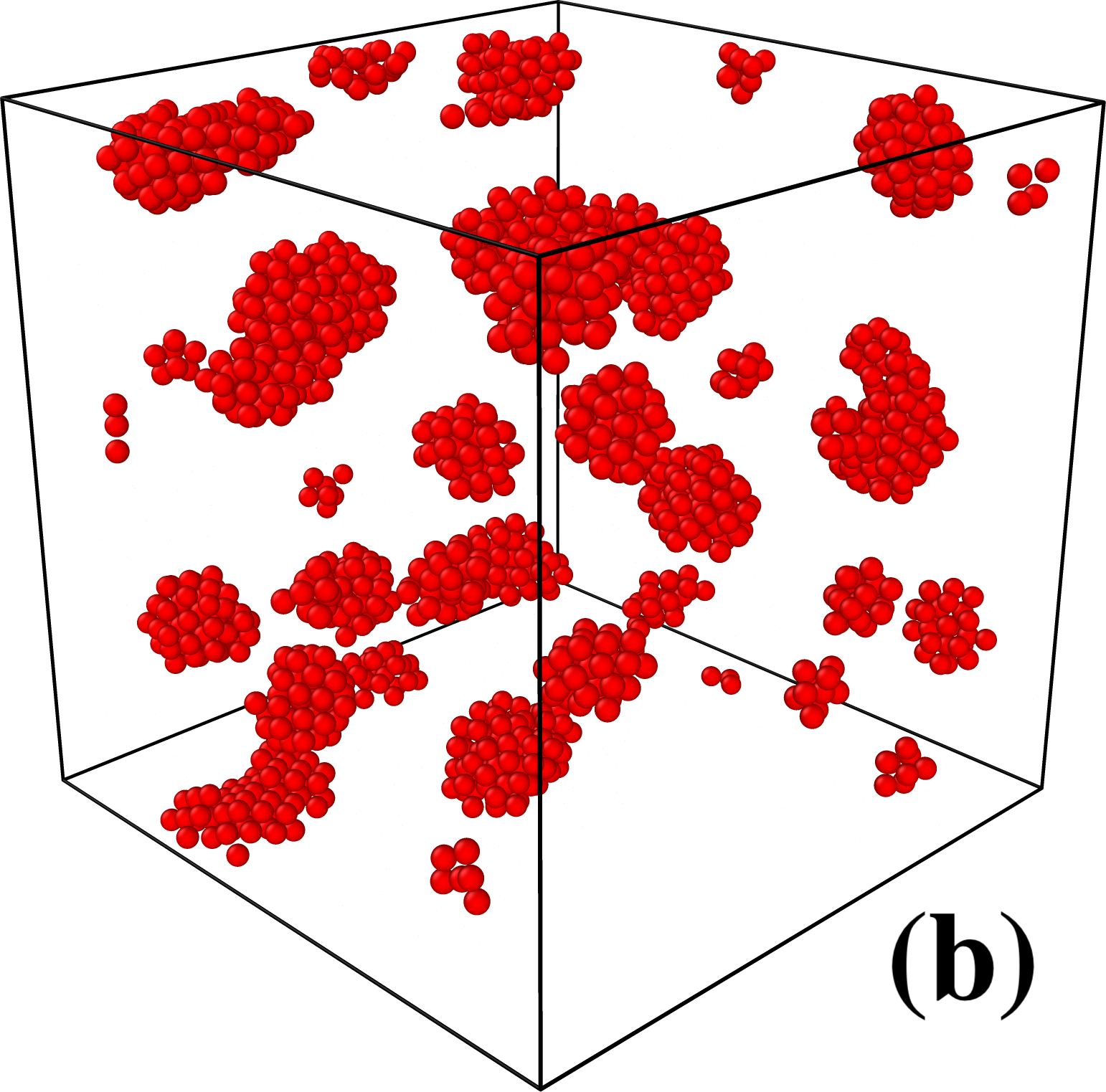}
 \end{subfigure}\\
 	\begin{subfigure}[h!]{.48\columnwidth}
 		\centering
 		\includegraphics[width=\columnwidth]{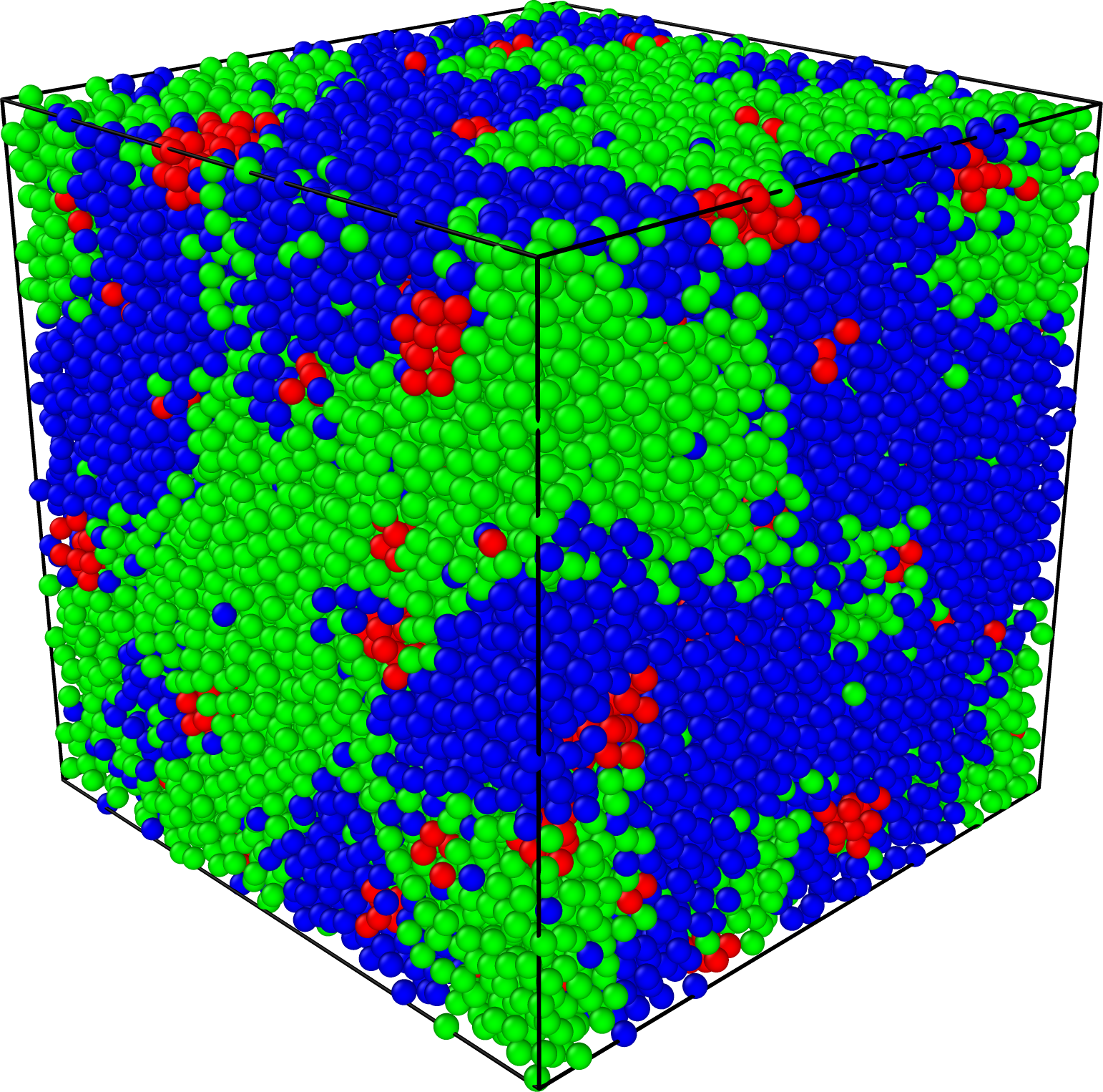}
 	\end{subfigure}
 \begin{subfigure}[h!]{.48\columnwidth}
 	\centering
 	\includegraphics[width=\columnwidth]{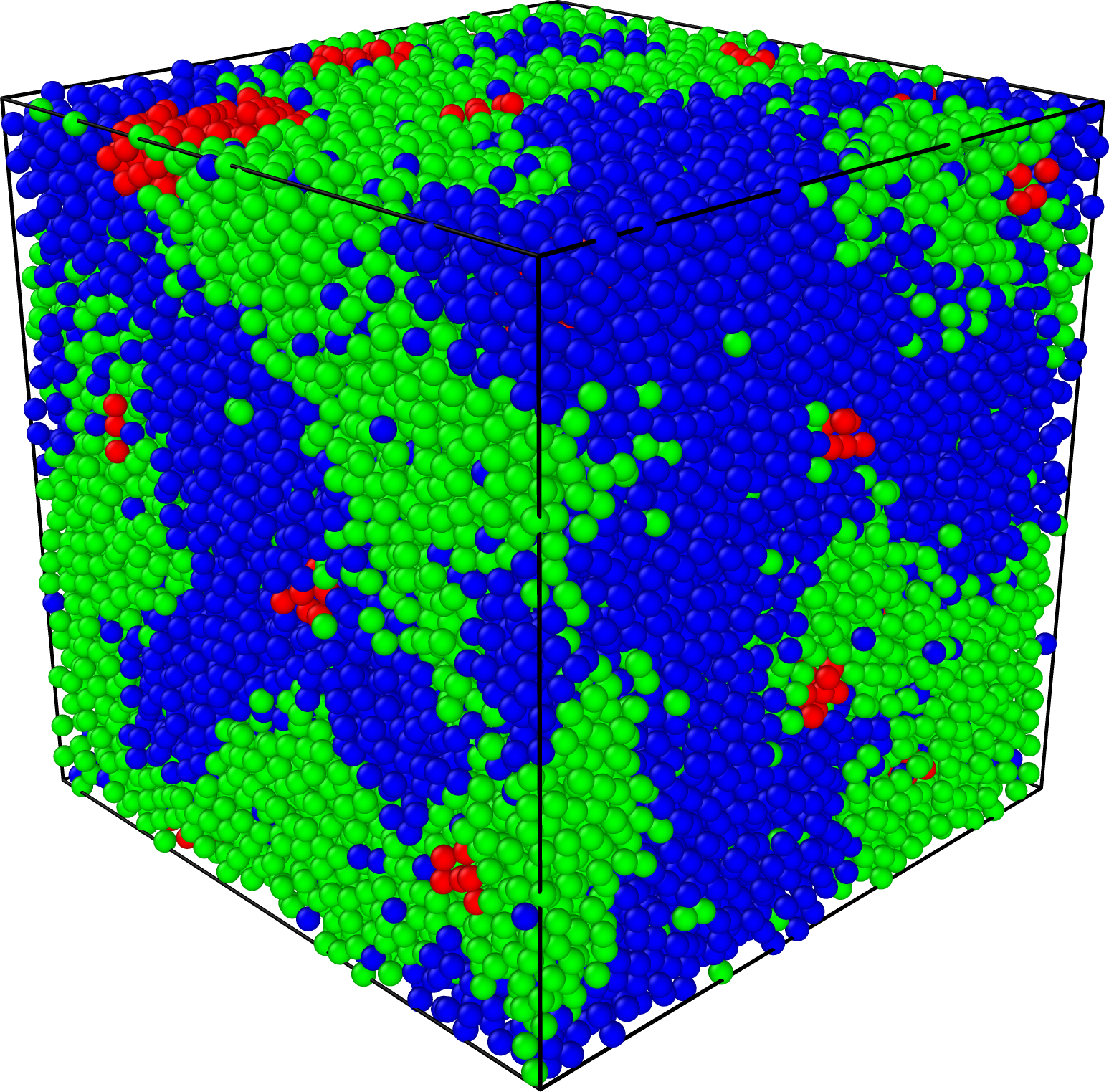}
 \end{subfigure}
 	\caption{The spacial distribution of the disordered particles (P type) for $\tau = 200$ and 500 are shown in (a) and (b) respectively for model \textbf{II}. In the lower panel we show the complete configuration of our system at $t=3000$. The A, B and P type particles are marked as green, blue and red color respectively.}
 	\label{fig2-snap}
 \end{figure}

The Fig.~\ref{fig1-snap}, the snapshots illustrate the phase separation dynamics in our binary fluid system with quenched disorder in model \textbf{I}. We present the three-dimensional snapshot and the two-dimensional cross-sectional view of the same, for the pure system $(\rho_P = 0.0)$ and including pinned particles with $\rho_P = 0.05$ at time $t = 3000$. For both the cases, bicontinuous domains of A and B particles are formed. It is evident from Fig. \ref{fig1-snap} that the size of the same species domains is smaller in the presence of disorder at a given time $t=3000$. Therefore, the domain growth slows down in the presence of quenched disorder. In model \textbf{II}, the spacial distribution of the disorder particles are shown in Fig.~\ref{fig2-snap} for the chosen value of $\tau=50, 200$ and $800$. As expected, the average cluster size for the $P$ type of particles increases with increase in $\tau$. In the same figure we also show the representative snapshots of the whole system at time $t=3000$. It is conspicuous that the sluggishness of the growth dynamics is relaxed for higher $\tau$ value. The aforementioned observation regarding the slowing down of the domain growth dynamics will be further quantified and elaborated upon in terms of the domain length scale $\ell(t)$ at a later stage.

The slower rate of domain growth can be elucidated as follows. During the process of segregation, particles of same species exhibit an attraction towards each other, leading to the formation of domains that evolve over time. Since the interaction between different species is weaker, the boundaries between domains become energetically favorable positions for the pinned particles. The domains of both the liquids organize themselves in a manner that ensures the pinned particles are positioned along the interfaces. Upon careful visual analysis of our simulation data, we have observed that the majority of the impurity P type particles consistently reside along these interfaces throughout the simulation. This assertion is supported by the snapshot depicted in Fig.\ref{fig1-snap} and \ref{fig2-snap}. Consequently, the pinned particles act as traps for the domain wall boundaries at these favorable positions. However, this effect is eventually overcome due to the presence of sufficiently high thermal energy. Additionally, we find that the domain boundaries become rough in the presence of quenched disorder.

To characterize the domain size and interfaces in the presence of quenched disorder, we resort to the spherically averaged so-called two-point equal time correlation function $C_{\psi\psi}(r,t)$:
\begin{equation}\label{Correlation_function}
	C_{\psi\psi}(r,t) = \langle\psi(0,t)\psi(\vec{r},t)\rangle /\langle\psi(0,t)\rangle^2
\end{equation} 
 where $\psi(\vec{r},t)$ is the order parameter. The later is computed based on the local density difference between the A and B particles. For that, the entire simulation box is divided into small cubic cells of size $(2\sigma)^3$ and the density difference $\delta\rho(\vec{r},t)= \rho_A(\vec{r},t)-\rho_B(\vec{r},t)$ is calculated for each cell. For  $\delta\rho(\vec{r},t)>0$, $\psi(\vec{r},t)$  takes up the value $+1$, and otherwise $-1$ . The angular brackets in Eq. \ref{Correlation_function} denote statistical averaging. The fourier transform of $C_{\psi\psi}(\vec{r},t)$ gives us the structure factor as $S(\vec{k},t) = \int d\vec{r} \hspace{0.1cm} exp(i\vec{k}.\vec{r}) \hspace{0.1cm} C_{\psi\psi}(\vec{r},t)$. 

\begin{figure}
\centering
\includegraphics[width=0.8\columnwidth]{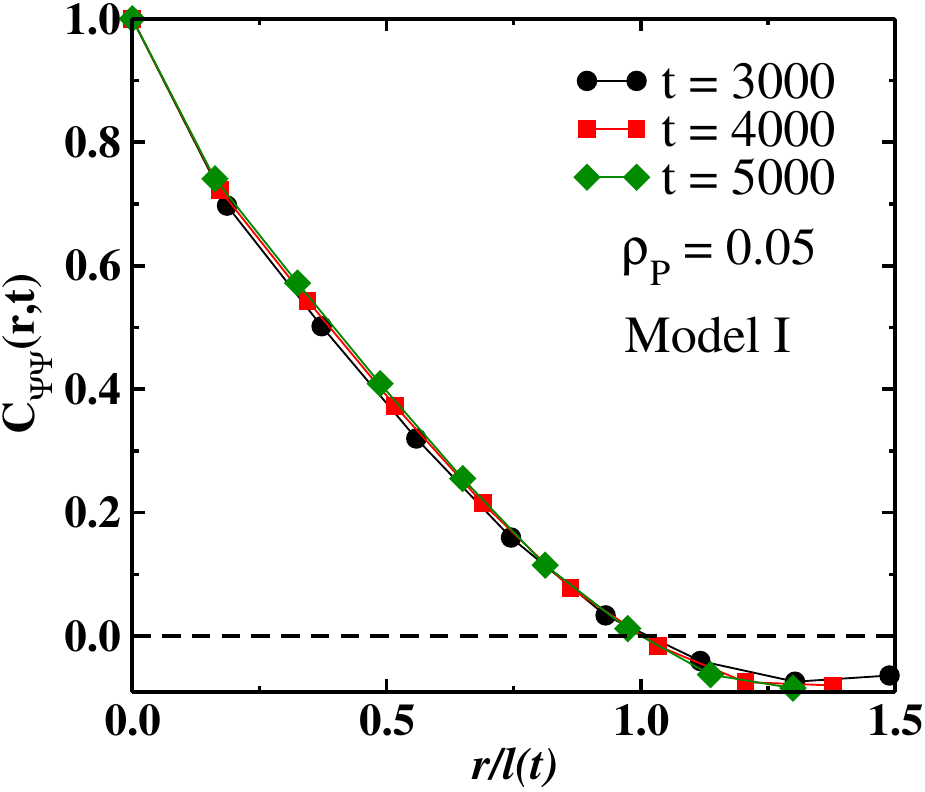}
\caption{The scaling plot of $C_{\psi\psi}(r,t)$ vs $r/\ell(t)$.}
\label{fig3}
\end{figure}

\begin{figure}
	\centering
	\begin{subfigure}[h!]{0.8\columnwidth}
		\includegraphics[width=\columnwidth]{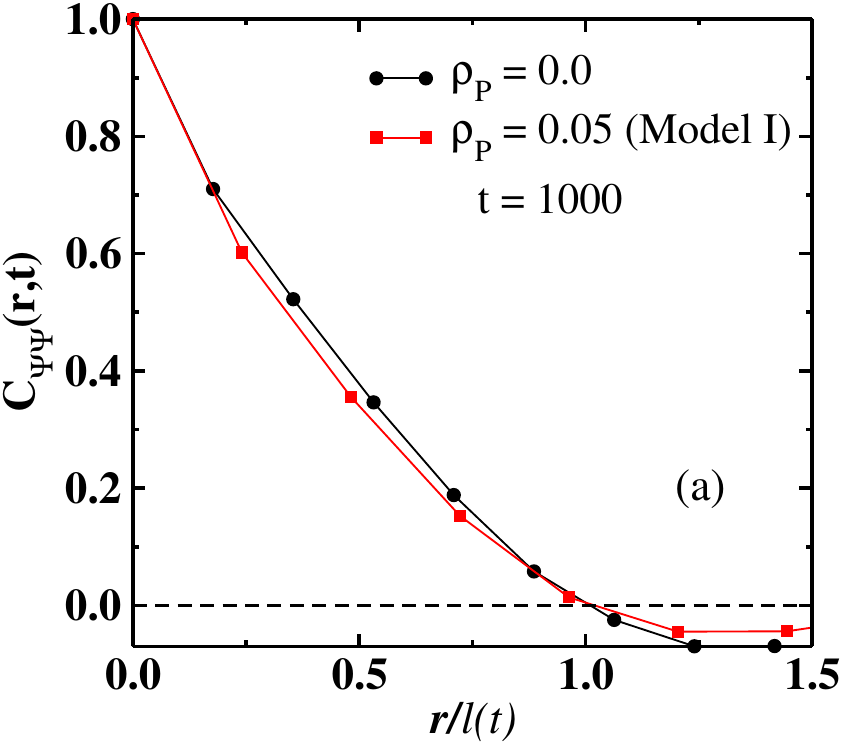}
	\end{subfigure}\\
	\begin{subfigure}[h!]{0.8\columnwidth}
		\centering
		\includegraphics[width=\columnwidth]{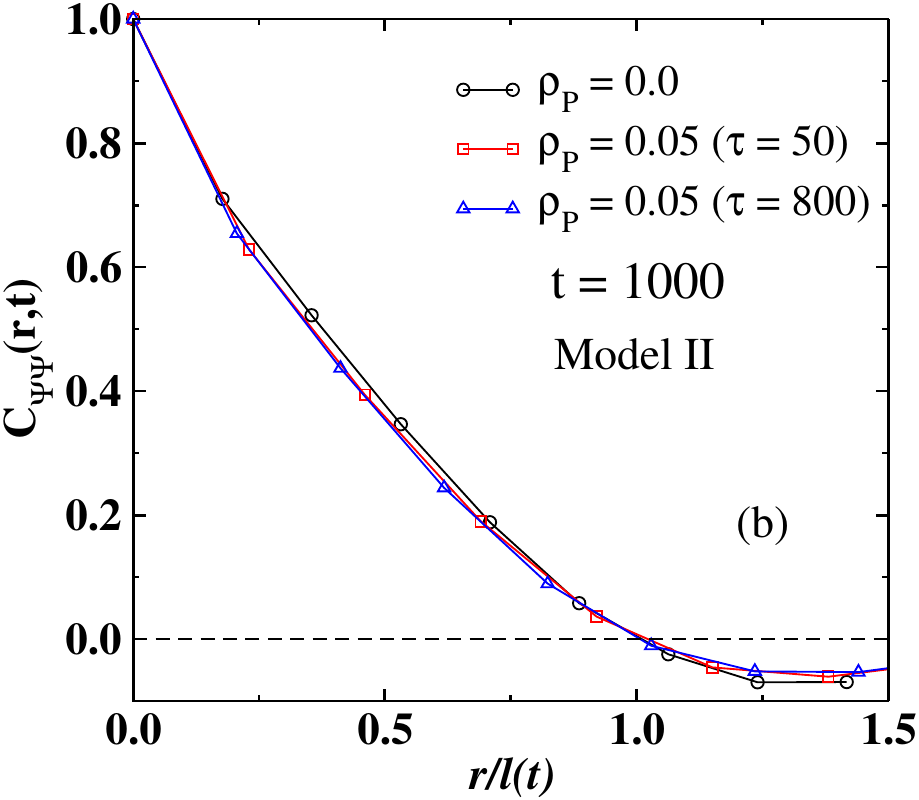}
	\end{subfigure}
	\caption{The scaling plot of $C_{\psi\psi}(r,t)$ vs $r/\ell(t)$ for (a) model \textbf{I} and (b) model \textbf{II} at time $t=1000$.}
	\label{fig4}
\end{figure}

In Fig. \ref{fig3}, we show the scaling plot of the correlation function $C(r,t)$ vs $r/\ell(t)$ for the system with disorder concentration $\rho_P = 0.05$. Here $\ell(t)$ is the average domain size, measured through the first zero of $C(r,t)$ in model \textbf{I}. We see a neat data collapse for different times. This observation validates that the system remains within the same universality class even in the presence of disorder \cite{Binder3}. We observe a similar scaling behavior for the other impurity concentration also (not shown). However, the correlation function gets modified when the impurity concentration $\rho_P$ changes. This is demonstrated in Fig. \ref{fig4}a, where we plot the $C(r,t)$ vs $r/\ell(t)$ for $\rho_P=0.0$ and 0.05 at a specific time $t$. For the sake of clear visualization, all the $\rho_P$ values chosen in our simulation are not shown in the same figure. Clearly, the different $C(r,t)$  do not overlap with one another. In the case of the pure system, the $C(r,t)$ exhibits a linear behavior at small values of $r$, commonly known as the Porod law. However, in systems with quenched disorder a non linear cusp is observed indicating a non Porod behaviour \cite{Gaurav}. These results are indicative of the roughening of domain boundaries in the presence of quenched disorder. A similar behavior is observed in model \textbf{II}, shown in Fig. \ref{fig4}b. Therefore, we conclude that the system with quenched disorder violets the Porod law \cite{Shaista,Saikat,RBBSG1}.

\begin{figure}
	\centering
	\includegraphics[width=0.8\columnwidth]{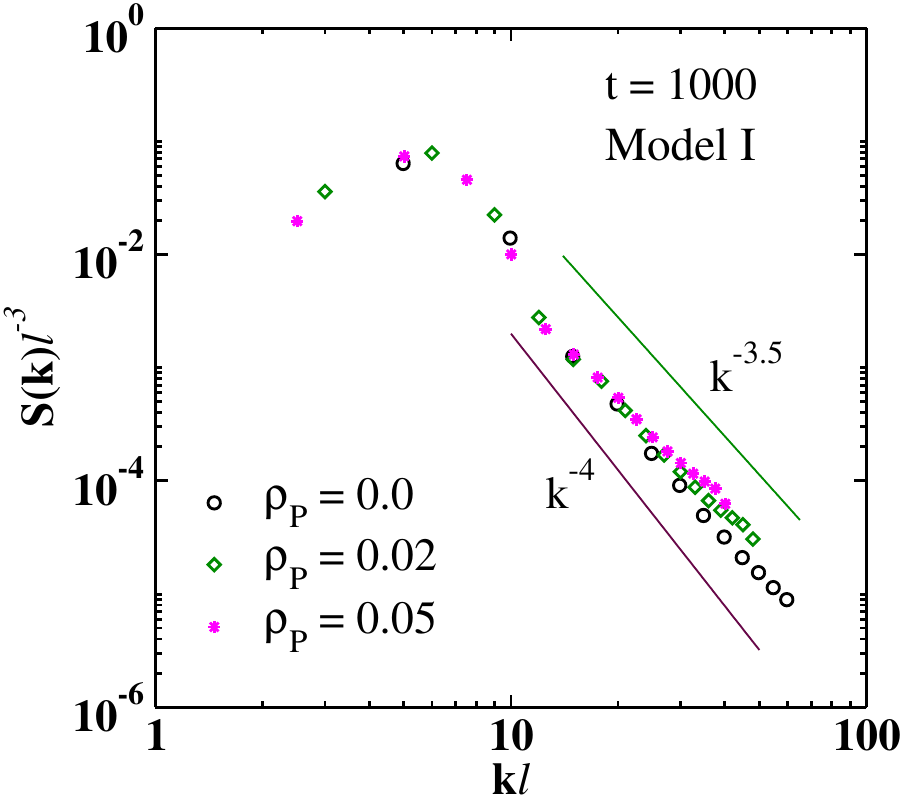}
	\caption{The scaled structure factor $S(k) \ell^{-3}$ vs $\ell k$ for different $\rho_P$ values in the log-log plot.}
	\label{fig5}
\end{figure}

To gain further insight into the morphology of the domain boundaries, we calculate the structure factor $S(\vec{k},t)$, This is shown in Fig. \ref{fig5}. As expected, for the pure system, the tail part of $S(\vec{k},t)$ follows the Porod law $S(k,t ) \sim k^{-(d+1)}$ \cite{Binder3}.  This is indicative of compact domains with sharp interfaces. In the presence of disorder, the large-$k$ behavior of $S(\vec{k},t)$ is modified and deviates from the Porod law. This can be attributed to the  roughening of the domain interfaces due to quenched disorder. This result reconfirms the violation of Porod law, as was demonstrated earlier in terms of correlation function \cite{Gaurav}. 

\begin{figure}
	\centering
	\begin{subfigure}[h!]{0.8\columnwidth}
		\includegraphics[width=\columnwidth]{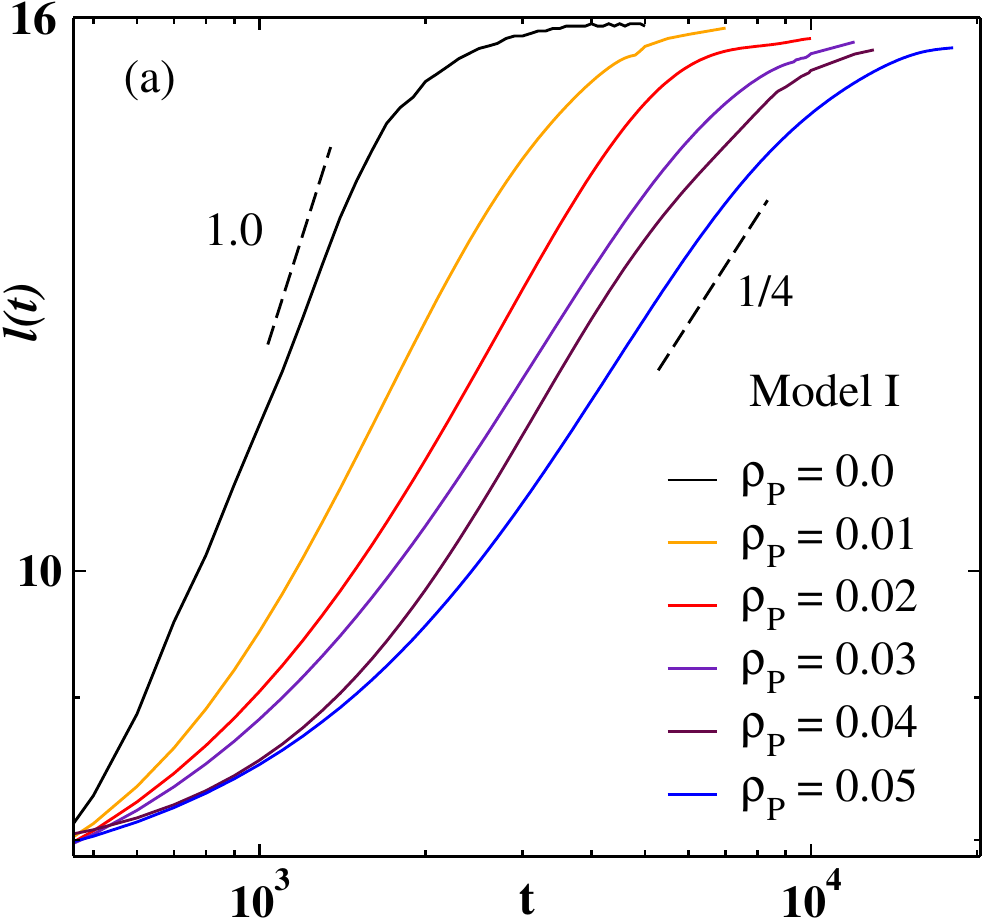}
	\end{subfigure}\\
	\begin{subfigure}[h!]{0.8\columnwidth}
		\centering
		\includegraphics[width=\columnwidth]{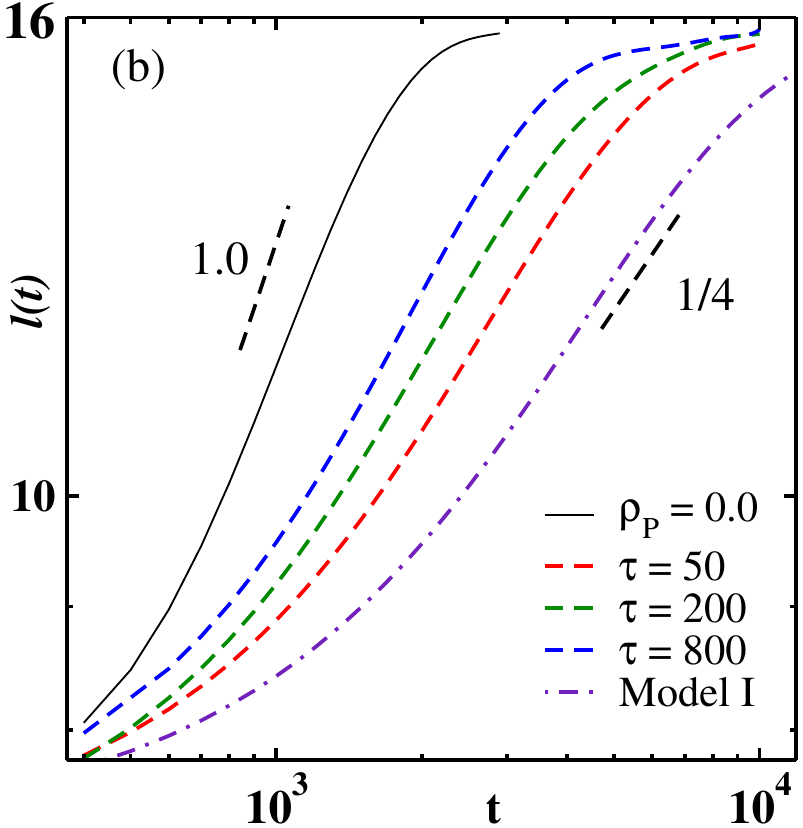}
	\end{subfigure}
	\caption{(a) The time evolution of the average domain size $\ell(t)$ is shown for different $\rho_P$ in the log-log scale for model \textbf{I}. (b) The same result is shown for model \textbf{II} for a particular value of $\rho_P = 0.05$ with dashed lines. We also show the $\ell(t)$ for the pure system (solid line) and for model \textbf{I} with $\rho_P = 0.05$ (dashed dotted line) for comparison.}
	\label{fig6}
\end{figure}

To understand the coarsening dynamics and domain size evolution, we focus on $\ell(t)$. In Fig. \ref{fig6}a, we show the $\ell(t)$ with time for model \textbf{I} for five different disorder densities $\rho = 0.01, 0.02, 0.03, 0.04$ and 0.05. After an initial transient period, our simulation successfully enters the viscous hydrodynamic regime. For pure system ($\rho_P = 0.0$), the $\ell(t)$ is expected to grow linearly with time \cite{Ahmad, Ahmad2012, Shaista}. Nonetheless, a non-zero offset at the crossover point is responsible for the minor deviation. By subtracting this offset from the lengthscale, we can restore the desired linear behavior. A dramatic slowing down in the domain growth is observed for disordered system and consequently the saturation time increases significantly with $\rho_P$. The observed slowest growth rate in our system corresponds to $\rho_P = 0.05$ and $\alpha \sim 1/4$. Note that, for the choice of higher impurity concentration ($\rho_P > 0.05$), the growth will further slow down accordingly.  In Fig. \ref{fig6}b we show the lengthscale $\ell(t)$ obtained using the model \textbf{II} with $\rho_P = 0.05$ for three different values of $\tau=50, 200$ and 800. As a reference, we also show the $\ell(l)$ for the pure system and in presence of disorder with $\rho_P = 0.05$ using model \textbf{I}. The slowing down of the coarsening dynamics is evident with decreasing $\tau$ value. These results are consistent with the observations shown in Fig. \ref{fig1-snap} and Fig. \ref{fig2-snap}.

Fig.  \ref{fig6}b illustrates a comparative study of the effect of disorder on the dynamics of the system for the two different models used in this work. We find that the effect of disorder decreases with increasing $\tau$ for model \textbf{II}. This can be understood as follows. As previously discussed, the disorder particles reside along the domain interfaces most of the time during the coarsening process. Therefore, these immobile particles serve as locations that trap the interfaces, causing a hindrance to the expansion of the domains.  As $\tau$ increases, the average size of clusters containing P particles grows, leading to a reduction in the overall number of such clusters. Consequently, the effective number of trapping sites decreases with $\tau$, and the dynamics becomes faster. Note that, model \textbf{I} provides the highest number of trapping sites along the interface, where the P type particle clustering is absent, and the motion is consequently the slowest.

\subsection{Aging dynamics}
\begin{figure}
	\centering
	\includegraphics[width=0.8\columnwidth]{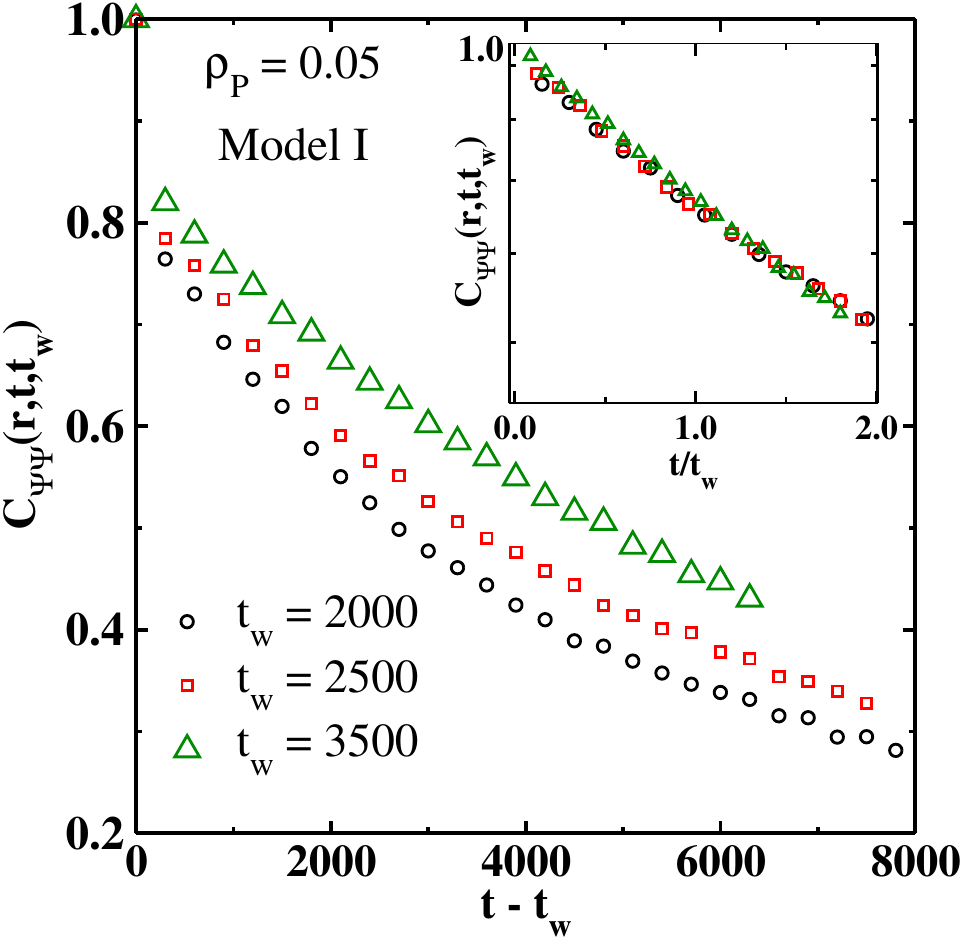}
	\caption{The plot of $C(r,t,t_w)$ vs $t - t_w$ for $(\rho_P = 0.05)$ at different $t_w$ for model \textbf{I} is shown in the main figure. In the inset we show the same data after rescaling the time with $t_w$. }
	\label{fig7}
\end{figure}

One of the hallmark of the non-equilibrium systems is the aging phenomena. When our binary system is driven out of equilibrium via rapid cooling, it shows aging behavior characterized by a lack of time-translational invariance. The aging dynamics in disorder free system is a well studied subject and we refrain from presenting those results \cite{Ahmad2012}. On the other hand, this topic is completely unexplored for off lattice systems with quench disorder. As mentioned in the introduction, to study the effect of such disorder on the aging dynamics in the phase separating systems, we introduce the spherically averaged two-time correlation function $C_{\psi\psi}(r,t,t_w)$ as follows:
\begin{equation}\label{autoCorrelation_function}
	C_{\psi\psi}(r,t,t_w) = \langle\psi(\vec{r},t)\psi(\vec{r},t_w)\rangle - \langle\psi(\vec{r},t)\rangle\langle\psi(\vec{r},t_w)\rangle   
\end{equation}
Attention is paid to the viscous hydrodynamics regime where the influence of disorder is the most. Therefore, following  Fig. \ref{fig6} we select the appropriate values of $t_w$. In Fig. \ref{fig7} the $C_{\psi\psi}(r,t,t_w)$ vs. $t - t_w$ is depicted for our system containing maximum impurity of $\rho_P=0.05$ in model \textbf{I}. Evidently, the correlation curves associated with different $t_w$ exhibit no overlap, indicating a clear violation of time translation invariance. In accordance with Fisher and Huse, we endeavor to scale the x-axis with $t/t_w$, resulting in a notable data collapse, as demonstrated in the inset of Fig. \ref{fig7}. We repeated the same exercise for the other impurity concentrations and model \textbf{II} (not shown) and observed the same behavior. Thus, the validity of the Fisher and Huse scaling law is reaffirmed for systems including quenched disorder \cite{FisherHuse}.

\begin{figure}
	\centering
	\includegraphics[width=0.8\columnwidth]{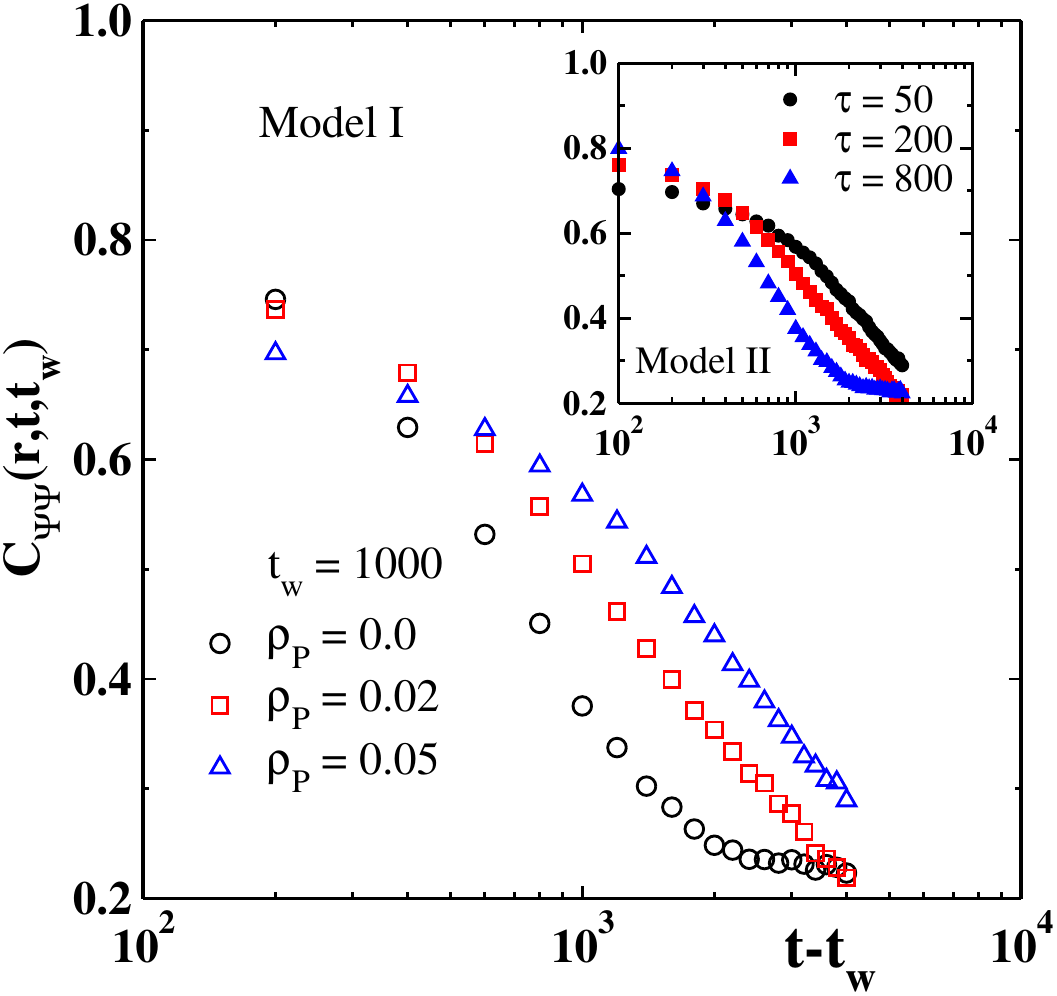}
	\caption{The $C(t,t_w)$ vs $t - t_w$ plot at $t_w=1000$ for three different impurity concentrations for model \textbf{I} is shown in the main figure. In the inset we show the same graphs for model \textbf{II} at $\rho_P=0.05$. }
	\label{fig8}
\end{figure}

To further explore the impact of impurities on aging dynamics, we calculate the $C_{\psi\psi}(r,t,t_w)$ for the pure and disordered system at a fixed $t_w = 1000$. This is shown in Fig. \ref{fig8} for both the models. Evidently, as the concentration of impurities increases, the correlation function exhibits a noticeably slower decay. This observation demonstrates the deceleration in the aging dynamics, which aligns with the findings depicted in Fig. \ref{fig6}.

\begin{figure}
	\centering
	\includegraphics[width=0.8\columnwidth]{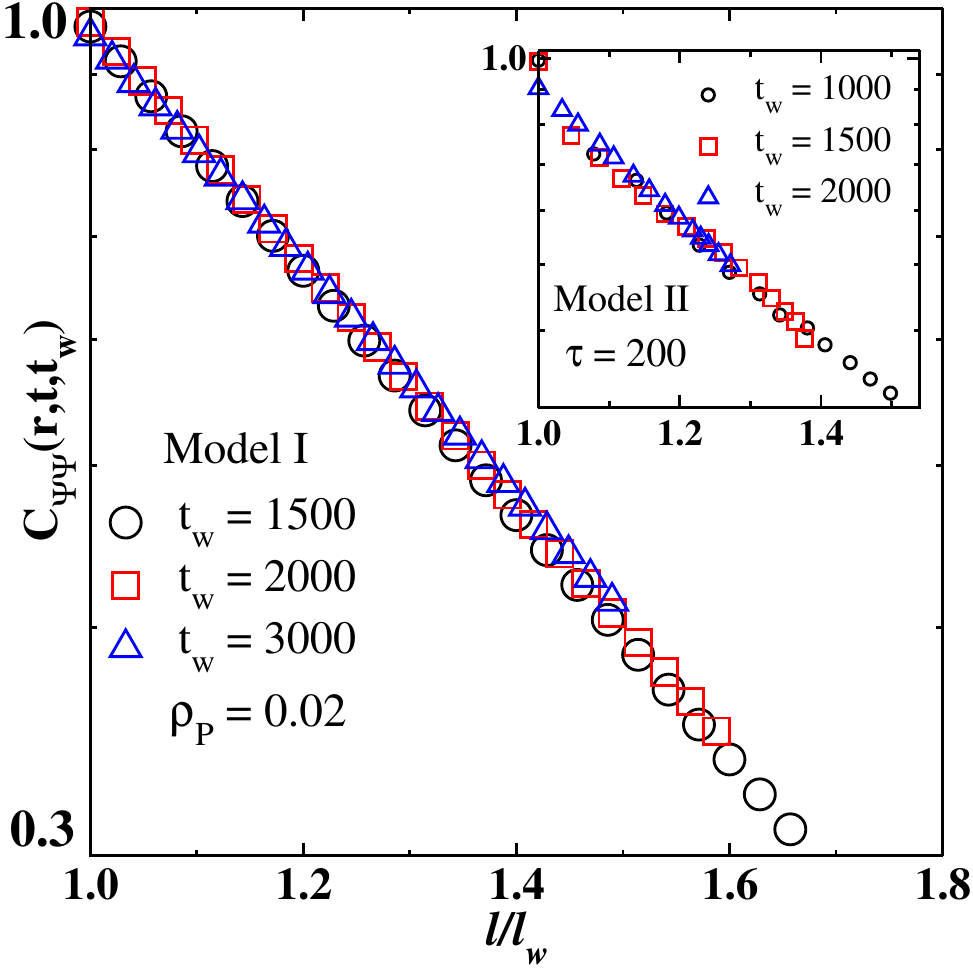}
	\caption{The $C(t, t_w )$ plotted as a function of $\ell/\ell_w$ for $\rho_P = 0.02$ at different $t_w$ values in the semi-log scale in the main figure. In the inset we show the same curves for model \textbf{II}.}
	\label{fig9}
\end{figure}

In Fig. \ref{fig9} we plot the $C_{\psi\psi}(r,t,t_w)$ with $\ell/\ell_w$ in a semi-log scale at a chosen impurity concentration $\rho_P=0.02$. A nice collapse of data is found for both the models \cite{Ahmad2012}. The data set exhibits a linear trend, providing confirmation of its exponential nature. Note that, the same exponential behavior is observed in the pure system also. Hence, we can conclude that the scaling law in the hydrodynamic regime is highly universal, oblivious to the presence of quenched disorder. This particular nature of decay is related to the advective hydrodynamic flows under the hydrodynamic effects \cite{Ahmad2012}.

\section{Summary and Discussion}
In summary, we have investigated the effect of quenched disorder on the kinetics of phase separation in a binary liquid system using molecular dynamics simulations. The disorder was introduced by incorporating a small concentration of immobile particles in the system. We used two different models to characterize the spacial distribution of the disorder particles.  The main observation is the substantial deceleration of domain growth dynamics. The disorder particles were found to be located at the energetically favorable domain interfaces. As a result, the domain boundaries were trapped and become rough. This led to the modified spacial correlation function and the violation of Porod law. The same effect was observed in the structure factor at the large $k$ limit, confirming the Porod law violation. For disordered system, the growth dynamics was studied by computing the characteristic length scale $\ell(t)$ with time. We found that, in such systems, the algebraic domain growth persists, with a decrease in the exponent as the impurity concentration is increased. 

Another important topic, studied in such nonequilibrium system is the aging phenomena. For the aging related studies we used the standard tool namely the two-time order-parameter autocorrelation function $C_{\psi\psi}(r,t,t_w)$. A clear violation of time translational invariance was observed. We demonstrated that the $C_{\psi\psi}(r,t,t_w)$ adheres to the scaling law proposed by Fisher and Huse with respect to $t/t_w$ in the disordered system. However, the aging process experienced a significant deceleration, with increasing impurity concentration. Additionally, we investigated the scaling law with respect to $\ell/\ell_w$ and obtained an excellent data collapse for all the cases under consideration. In the semi log scale, the scaling function exhibited a linear decay indicating an exponential nature in the hydrodynamic regime. Consequently, we conclude that the scaling laws governing aging dynamics are generic and universal, even in the presence of disorder.\\

\noindent{\it Acknowledgement.---} B. Sen Gupta acknowledges Science and Engineering Research Board (SERB), Department of Science and Technology (DST), Government of India (no. CRG/2022/009343) for financial support. R. Bhattacharyya acknowledges VIT for doctoral fellowship.

\end{document}